\documentclass[twocolumn,showpacs,preprintnumbers,amsmath,amssymb]{revtex4}

\usepackage{graphicx}
\usepackage{dcolumn}
\usepackage{bm}
\def\prd{Phys.\ Rev.\ D}
\def\prl{Phys.\ Rev.\ Lett.}
\def\plb{Phys.\ Lett.\ {\bf B}}
\def\npb{Nucl.\ Phys.\ {\bf B}}

\begin{document}

\preprint{BIHEP-TH-2003-36}

\title{The Color Octet Effect from $e^+ e^-\rightarrow{J/\psi}+X+\gamma$ at B Factory} 

\author{Jian-Xiong Wang}
\email{jxwang@mail.ihep.ac.cn}
\affiliation{ Institute of High Energy Physics, Academia Sinica, Beijing 100039, CHINA}
 \homepage{http://www.ihep.ac.cn/lunwen/wjx/public_html/index.html}

\date{\today}

\begin{abstract}
We study the initial state radiation process $e^+ e^-\rightarrow{J/\psi}+X+\gamma$ 
for $J/\psi$ production at B factory, and find the cross section
is $61\%$ larger than it's Born one 
for color octet part and is about half as it's Born one for color singlet part. 
Furthermore, the color singlet and color octet signal are very clearly separated
in it's $E_\gamma$ spectra due to kinematics difference. We suggest to measure
this $E_\gamma$ spectra at B factory to determine the color octet effect.  
\end{abstract}

\maketitle

To study $J/\psi$ production on various experiments is a very interesting topic since 
its discovery in 1974. It is a good place to probe of both perturbative and 
nonperturbative aspects of QCD dynamics. To describe the huge discrepancy of 
the high-$p_T$ $J/\psi$ production between the theoretical calculation based 
on color singlet mechanism\cite{j.h.kuhn:79} and 
the experimental measurement by the CDF collaboration at the Tevatron\cite{cdf1}, 
color-octet mechanism\cite{fleming} was proposed based on 
non-relativistic QCD(NRQCD)\cite{nrqcd}. The factorization formalism of NRQCD
provides a theoretical framework to the treatment of heavy-quarkonium production. 
The color singlet mechanism is straightforward from the perturbative QCD, 
but the color-octet mechanism depends on  nonperturbative universal 
NRQCD  matrix elements. So various efforts have been made to confirm this 
mechanism, or to fix the magnitudes of the universal NRQCD matrix elements. 
Although it seems to show qualitative agreements with experimental data, there
are certain difficulties in the quantitative estimate of the
color-octet contribution in $J/\psi$ and $\psi'$ photoproduction at HERA
\cite{cacciari:yr96,Amundson:yr97,ko:yr96,Kniehl:yr97,kramer:yr96},
$J/\psi(\psi')$ polarization in large transverse momentum production at
the Fermilab Tevatron \cite{beneke:96yr,braaten:99yr,leibovich:97yr},
and more recently in B-factories.
  
There are at least two kinds of uncertainty in the theoretical treatment of 
$J/\psi$ production.  One is from the QCD correction. It is difficulty to estimate
the QCD correction effect up to a few times in various leading order calculation  
for $J/\psi$ production. The uncertainty to fix the NRQCD matrix elements 
is quit large for the case in which color singlet effect is compatible with the 
color octet one such as in the measurement of the $J/\psi$ production at B factory in BaBar
and Belle experiments\cite{BABAR1,BELLE1}. There are a few examples shown that the NLO 
correction are quit large.  
It was found that the current experimental results on 
inelastic $J/\psi$ phtonproduction\cite{hera:h1,hera:zeus} are adequately 
described by the color singlet channel alone once higher-order QCD corrections 
are included\cite{kramer:yr96}.
Although ref\cite{klasen:yr02} found that the new DELPHI
\cite{delphi:yr01} data evidently favor the NRQCD formalism for
$J/\psi$ production $\gamma + \gamma \rightarrow J/\psi + X$, but rather the 
color-singlet model. It was also found 
by ref\cite{qiaow0301} that 
the QCD higher order process $\gamma + \gamma \rightarrow J/\psi c \bar{c}$ 
gives the same order and even bigger contribution at large $p_T$ compare 
to leading order color singlet processes.
ref\cite{qcf:03} found that higher order process give large contributions. 

\begin{figure}
\includegraphics{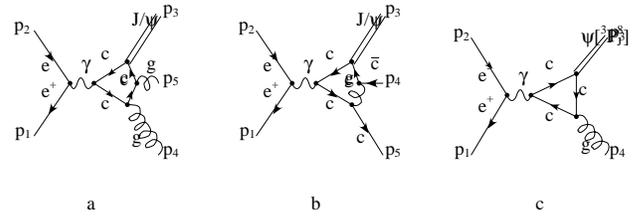}
\caption{\label{fig:d1}The typical Feynman diagrams for the $e^+  e^- \rightarrow J/\psi X$,  
there are 6 diagrams for the $e^+ e^- \rightarrow {J/\psi} gg$, 
4 diagrams for $e^+ e^- \rightarrow {J/\psi}c\bar{c}$ and  
2 diagrams for each $e^+ e^- \rightarrow {\psi[^3P^8_J]} g,~J=0,1,2$}
\end{figure}

The another uncertainty is from the hadronization of the color octet $J/\psi^{(8)}$ 
to the color singlet $J/\psi$ and hadrons.  
The recent measurement of the $J/\psi$ production at B factory in BaBar 
and Belle experiments\cite{BABAR1,BELLE1} show that the theoretical predication
\cite{braaten:96,p.cho:96} for 
$p^*_{J/\psi}$ spectra did not agree with experimental results,   
and gives the total cross section
$\sigma(e^+e^-\rightarrow J/\psi+X)=1.47\pm0.10\pm0.13pb$.
It seems compatible with theoretical prediction\cite{f.yuan:97,s.baek:98,schuler:99} 
cover the range $0.8-1.7pb$
, in which about $0.3pb$ from color singlet processes 
$e^+e^-\rightarrow J/\psi gg$ and $e^+e^-\rightarrow J/\psi c\bar{c}$,
and $0.5-0.8pb$ from color octet processes $ e^+e^-\rightarrow J/\psi^{(8)}g$.  
The color octet processes only contribute to the endpoint of $p^*_{J/\psi}$ spectra
due to the kinematics of the two body final state.  But the experiments did not 
observe the this signal. To explain this discrepancy,
it is nature to think that $J/\psi^{(8)}$ have to hadronize into 
color singlet $J/\psi$ and will lose it's energy such as the case when a quark
jet hadronize into hadrons. Therefore there should be a hadronization possibility 
function $F(x,...)$ for $J/\psi$ production with momentum 
$p_{J/\psi}=x~p_{J/\psi^{(8)}}$ and $\int^1_0dx~F(x,...)=1$. 
The universal NRQCD matrix elements treatment is
just a first step approximation of $F(x,...)=\delta(1-x)$. 
The ref\cite{fleming2003} tried on it for the $J/\psi$ production at B factory
and broadened the $p^*$ spectra from color octet $J/\psi$.
\begin{figure}
\includegraphics{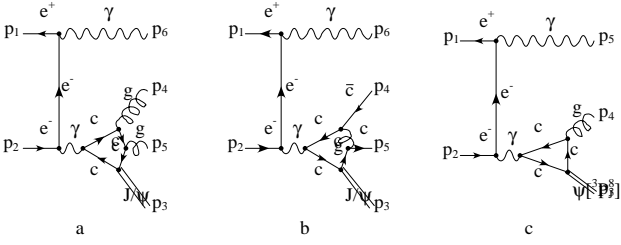}
\caption{\label{fig:d2}The typical Feynman diagrams for the $e^+  e^- \rightarrow 
J/\psi X\gamma$, there are 12 diagrams for the 
$e^+ e^- \rightarrow {J/\psi} gg\gamma$, 
8 diagrams for $e^+ e^- \rightarrow {J/\psi}c\bar{c}\gamma$ and  
4 diagrams for each $e^+ e^- \rightarrow {\psi[^3P^8_J]} g\gamma,~J=0,1,2$}
\end{figure}

Due to the above two kinds of uncertainty, the $p^*_{J/\psi}$ spectra 
from color octet $J/\psi^{(8)}$ is unknown, and meanwhile the higher order 
QCD effect for the color singlet $J/\psi$ production is also unknown. So 
to confirm, to extract 
$J/\psi^{(8)}$ information from the measurement in B factory experiment
become unavailable. However, we find a good way to determine  
the color singlet and color octet effect, i.e. to measure the $E_\gamma$ spectra
in the initial state radiation process 
$e^+ e^-\rightarrow{J/\psi}+X+\gamma$, which could avoid the uncertainties from 
$J/\psi^{(8)}$ hadronization and QCD correction.  

This QED higher order process is thought to be small at first glance. 
But there are already few examples in which QED higher order process could be
very lager in certain case.  
It was found that there are large contributions to $J/\psi$ production form QED 
processes such as $e^+ e^- \rightarrow J/\psi  \gamma$ and
$e^+ e^- \rightarrow J/\psi  e^+  e^-$  \cite{chang:yr97} due to enhancement
from the t-channel peak. In the try to explain the double $J/\psi$ production 
measured by Belle\cite{BELLE2},  ref\cite{braaten:03} 
found that the high order QED process  $e^+  e^- \rightarrow J/\psi+J/\psi$ 
gives large contribution. Our numerical results show 
that $e^+ e^-\rightarrow{J/\psi}X+\gamma$ gives larger cross section than
$e^+  e^- \rightarrow J/\psi + X$.

For the leading process $e^+  e^- \rightarrow J/\psi + X$,
the typical Feynman diagrams are shown in  
Fig.\ref{fig:d1}. 
The color singlet processes 
$e^+ e^- \rightarrow {J/\psi} g g,~{J/\psi} c \bar{c}$ 
and color octet processes 
$e^+ e^- \rightarrow {\psi[^3P^8_J]} g,~J=0,1,2;~~{\psi[^1S^8_0]} g$ 
were calculated in ref.\cite{braaten:96,p.cho:96,s.baek:98,f.yuan:97}. 
For the $e^+  e^- \rightarrow J/\psi + X+\gamma$,there are 
$e^+ e^- \rightarrow {J/\psi} g g \gamma,~{J/\psi} c \bar{c}+\gamma$, 
$~{\psi[^1S^8_0]} g {\gamma}, 
~{\psi[^3P^8_J]} g {\gamma},~J=0,1,2$, and 
the typical Feynman diagrams are shown in Fig.\ref{fig:d2}.  

To calculate this initial state QED radiation correction,  
we have to include two parts, One is the virtual correction plus soft photon emitted 
process, the another is the hard photon emitted process.
A universal formula is easily obtained for the first part, i.e. virtual correction  
plus soft photon emitted one. It is\cite{IRC}:  
\begin{eqnarray}\label{eq:1}
\sigma^{SV}=\sigma_0\{\frac{2 \alpha}{\pi}[(\ln{\frac{s}{m^2_e}}-1)
(\ln{\frac{2\delta\epsilon}{\sqrt{s}}}+\frac{3}{4})+\frac{\pi^2}{6}-\frac{1}{4}]\}.
\end{eqnarray} 
Where the $\sigma_0$ is the Born cross section, the $\delta\epsilon$ is maximum  
energy of the emitted soft photon, $s$ is 
the center mass energy of $e^+ e^-$ collider. In the above formula, the infrared
divergence are canceled between the soft photon emitted process and virtual 
correction.  

For both the Born processes and the hard photon emitted processes,
We obtain the formula and Fortran source of the cross section, 
including the kinematics, by using our automatic Feynman diagram calculation 
program FDC\cite{wang:fdc96}. The FDC was used in the calculation of 
$e^+ e^- \rightarrow J/\psi  \gamma$  and
$e^+ e^- \rightarrow J/\psi  e^+  e^-$  \cite{chang:yr97}. 
In ref\cite{qiaow0301} it was used in the calculation of  
$\gamma + \gamma \rightarrow J/\psi c \bar{c}$, and many other processes
to repeat the calculation by ref\cite{klasen:yr02}.
In the Fortran source generated by FDC, it is very easily 
to check the gauge invariance by replacing the polarization vector of gluon or 
$\gamma$ to it's momentum. All the possible gauge invariance are checked to be 
satisfied numerically in our calculation. Our analytic formula generated by FDC for       
$e^+ e^- \rightarrow {\psi[^3P^8_J]} g,~J=0,1,2;~~{\psi[^1S^8_0]} g$ are checked  
with ref\cite{braaten:96}.

\begin{table}
\caption{\label{tab:table1}The numerical results for the color singlet 
and color-octet $J/\psi$ production. 
The $p^*_{J/\psi}=\sqrt{p^2-m^2_{J/\psi}}$ is the 3-momentum of $J/\psi$.
$\delta\epsilon=0.1$ is used.
The cut $p^*_{J/\psi}>2.0GeV$ is used by the experiment 
to suppress the $J/\psi$ from $B$ decay. 
The result for color octet with cut
condition $p^*_{J/\psi}>2.0GeV$ is not presented since it does not make sense 
in the case that the detail of $J/\psi^{(8)}$ hadronization procedure is unknown} 
\begin{ruledtabular}
\begin{tabular}{lcccr}
 & Born & Virtual+Soft& hard photon \\
\hline
$p^*_{J/\psi}>2.0GeV$&&&&\\
\hline
$e^+ e^-\rightarrow{J/\psi}gg(\gamma)$ & 0.158 & -0.044 &0.072 \\
$e^+ e^-\rightarrow{J/\psi}c\bar{c}(\gamma)$ & 0.102 &-0.028 &0.022\\
In total  & 0.270 &-0.072 & 0.094 \\
\hline
$p^*_{J/\psi}>0GeV$&&&&\\
\hline
$e^+ e^-\rightarrow{J/\psi}gg(\gamma)$ & 0.203 & -0.056 &0.099 \\
$e^+ e^-\rightarrow{J/\psi}c\bar{c}(\gamma)$ & 0.119 &-0.033 &0.029\\
In total  & 0.322 & -0.089 &0.128 \\
\hline
$e^+ e^-\rightarrow{\psi[^3P^8_J]}g (\gamma)$&0.143&-0.039&0.351\\
$e^+ e^-\rightarrow{\psi[^1S^8_0]}g (\gamma)  $&0.133&-0.037&0.094\\
In total  &0.276 &-0.076&0.445\\
\end{tabular}
\end{ruledtabular}
\end{table} 

\begin{figure}
\includegraphics{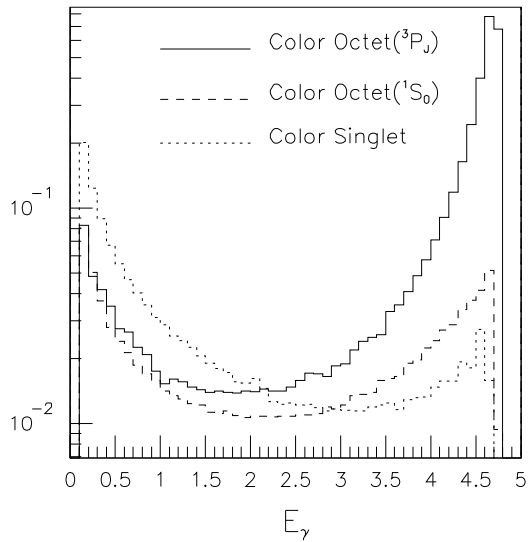}
\caption{\label{fig:p1}The differential cross section distribution vs the energy of 
emitted hard photon}
\end{figure}

\begin{figure}
\includegraphics{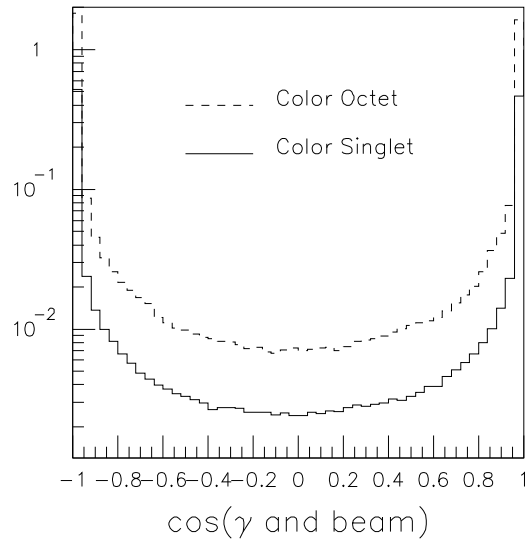}
\caption{\label{fig:p2}The differential cross section distribution vs the 
$cos(\theta)$ of the emitted hard photon and the beam}
\end{figure}

The numerical results are listed in Table.~\ref{tab:table1} and shown 
in Fig.\ref{fig:p1} and \ref{fig:p2}.  
In the numerical calculation, we use 
${m_{c}}=1.5$, ${m_{J/\psi}}=3.0$,
$\alpha=1/137$ and the lowest-order formula for 
$\alpha^{(n_f=3)}_s(\mu)$.
$J/\psi$ matrix elements are chosen as $<J/\psi>=1.4$. 
The universal NRQCD matrix elements for color octet $J/\psi$ are  
chosen as that in ref\cite{nrqcd},i.e, 
$<J/\psi^{(8)}(^3S_1)>=0.0039$, $<J/\psi^{(8)}(^1S_0)>=0.015$,
$<\psi^{(8)}(^3P_J)>=(2J+1) 0.0043 {m_{c}}^2 $. 

The renormalization scale $\mu$ dependent in the calculation 
are $\alpha_s=\alpha^{(3)}_s(10.6)=0.188$ for Born processes 
and $\alpha_s=\alpha^{(3)}_s(\sqrt{(p_1+p_2-p_\gamma)^2})$ 
for hard photon processes. 
Where the $p_1,p_2$ and $p_\gamma$ are the momentums for $e^+,e^-$ and the 
emitted $\gamma$ respectively. 
The hard photon processes are defined as $E_\gamma>\delta\epsilon$ 
, i.e. the $\delta\epsilon$ is the minimum energy of the emitted photon. 
We have used different value of $\delta\epsilon=0.5,0.4,0.3,0.2,0.1,0.01GeV$ 
to verify that the Eq.\ref{eq:1}, which 
included only the leading term ${1 \over k}$ from the soft photon, is a good 
approximation when $\delta\epsilon$ is small enough. The 
numerical results show that the result of virtual + soft + hard is almost
independent of $\delta\epsilon$. 
There is infrared divergence for the process
$e^+ e^-\rightarrow{\psi[^3P^8_J]}g (\gamma)$ when the energy of gluon approach 
to zero, but it can not happen since the gluon must hadronize into at least one pion. 
To apply the same cut condition used by the experiment\cite{BELLE1} 
to suppress the background, the condition $\sqrt{(p_1+p_2-p_\gamma)^2}>m_{J/\psi}+3m_{\pi}$ 
is applied to all the radiation processes. 

There are other processes which may give compatible contributions from
the perturbative order analysis, but the contributions form them are found too 
small so that they do not appeared in the above list. Such as 
$e^+ e^-\rightarrow{\psi[^3S^8_1]}g\gamma$ is $1.2\times 10^{-4}pb$. 
$e^+ e^-\rightarrow{J/\psi}gge^+ e^-$ with 48 Feynman diagrams are 
calculated and the result is $6.4\times10^{-4}pb$

In Fig.\ref{fig:p2}, we find that most of the emitted photon go out with small
angle to beam direction in both the color singlet and color octet processes. 
This is the t-channel enhancement effect which make the contribution of the QED 
correction large.  
In Fig.\ref{fig:p1}, it very clearly shown 
that the color singlet processes mainly emit the photon with 
low energy ($E_\gamma<1.0GeV$), meanwhile, the color octet processes mainly emit 
the photon with high energy ($E_\gamma>4.0GeV$). The reason for this distribution is 
completely form the kinematics structure as shown in Fig.\ref{fig:d2}. 
When the center mass energy go down from 10.6Gev to 5GeV, the cross section 
will increase about 4 times for 
$e^+ e^-\rightarrow{J/\psi}gg$, about 51 times for 
$e^+ e^-\rightarrow{\psi[^3P^8_J]}g$ and about 17 times for  
$e^+ e^-\rightarrow{\psi[^1S^0]}g$, and the 
channel $e^+ e^-\rightarrow{J/\psi}c\bar{c}$ will be closed. 
The above property was shown in the Fig. 5 in ref\cite{f.yuan:97} and 
the Fig. 3 in ref\cite{schuler:99}. 
So that the remaining subprocesses will show this property 
after the photon emitting from the initial electron or positron. 
The peak in the small $E_\gamma$ is well known from the soft photon effect. 
Based on the above analysis, it is very clear that the spectra in Fig.\ref{fig:p1}
comes from the kinematics structure of the hard photon emitted processes. 
It is hardly be changed even if there is large QCD correction effects since the
QCD correction appears only in the subprocesses and will not change the kinematics
structure. So the ratio of the cross section for hard photon process to that 
for it's Born processes will be hardly changed by the QCD correction. 
In another side, the $E_\gamma$ spectra can not be changed by the procedure of
$J/\psi^{(8)}$ and gluon hadronization into $J/\psi$ and hadrons
since the hadronization procedure has no relation with it.  

In the measurement of $J/\psi$ production at B factory in BaBar and 
Belle experiments\cite{BABAR1,BELLE1}, it was tried to separate the color
singlet and color octet effects by measuring the spectra of $p^*_{J/\psi}$. 
The spectra show that the theoretical predication for color octet does not 
appear. It let us have to think about the detail of how the $J/\psi^{(8)}$
go through hadronization procedure. The ref\cite{fleming2003} addressed
on the problem, but the results is not predictive since the introduced shape 
function for $J/\psi^{(8)}$ is unknown. To avoid the uncentainties from  
$J/\psi^{(8)}$ hadronization and QCD correction, we find that to measure 
the $E_\gamma$ spectra of the hard photon emitted $J/\psi$ production process 
$e^+ e^-\rightarrow{J/\psi}+X+\gamma$ in B factory experiment is a very 
clear and good way. This spectra is not changed by QCD correction, 
by the detail of $J/\psi^{(8)}$ hadronization. and will 
very clearly separate the color singlet and color octet signal. Another advantage
is that the color octet cross section in the hard photon process is even 
$61\%$ large than it's Born one.  

Furthermore, It has been argued in the above 
that the ratio of the cross section of the hard photon process to 
it's Born one will be hardly changed by QCD correction and the detail
of $J/\psi^{(8)}$ hadronization. As approximation, we have 
\begin{equation}\label{eq:2} 
\begin{array}{lc}
&\sigma(e^+ e^-\rightarrow{J/\psi}+X)=0.233 x + 0.2 y, \\
&\sigma(e^+ e^-\rightarrow{J/\psi}+X+\gamma)=0.128 x + 0.445 y,\\
&\sigma(e^+ e^-\rightarrow{J/\psi}+X,color~singlet)=0.233 x, \\
&\sigma(e^+ e^-\rightarrow{J/\psi}+X,color~octet)=0.2 y.
\end{array}
\end{equation} 
where $x$ and $y$ represent the effects of QCD correction and uncertainty from  
the NRQCD matrix elements, and all the numerical
values from the Table.\ref{tab:table1}.  The color octet and 
color singlet contributions can be extracted by solving Eq.\ref{eq:2} when 
both the cross section $\sigma(e^+ e^-\rightarrow{J/\psi}+X)$ and 
$\sigma(e^+ e^-\rightarrow{J/\psi}+X+\gamma)$ be measured in the B factory 
experiment. In this way, the color octet cross section could not apply
cut condition $p^*_{J/\psi}>2.0GeV$ since the $p^*_{J/\psi}$ spectra  
depends on $J/\psi^{(8)}$ hadronization. 
So how to suppress the background from $B$ decay 
becomes a problem. However the measurement of the $E_\gamma$ spectra will not 
suffer from this problem and the $E_\gamma$ spectra fitting will fix the 
contributions from color singlet and color octet mechanism.   
\begin{acknowledgments}
This work was supported in part by the National Natural Science foundation of China
under Grant Nos.90103013 and by the Chinese Academy of Sciences under Project 
No. KJCX2-SW-N02.
\end{acknowledgments}


\begin{thebibliography}{99}
\bibitem{j.h.kuhn:79} 
C.H. Chang, \npb{\bf 172} (1980) 425;\\
J.H. K\"uhn, J. Kaplan, and E.G.O. Safiani, 
\npb{\bf 157} (1979) 125;\\
B. Guberina, J.H. K\"uhn, R.D. Peccei, and R. R\"uckl,
\npb{\bf 174} (1980) 317;\\
E.L. Berger and D. Jones, \prd{\bf 23} (1981) 1521;\\
R. Baier and R. R\"uckl, Z. Phys. C{\bf 19} (1983) 251.

\bibitem{cdf1}
F.~Abe {\it et al.}  [CDF Collaboration],
Phys.\ Rev.\ Lett.\  {\bf 69} (1992) 3704.
Phys.\ Rev.\ Lett.\  {\bf 79} (1997) 572.
Phys.\ Rev.\ Lett.\  {\bf 79} (1997) 578.

\bibitem{fleming} 
E.~Braaten and S.~Fleming,
Phys.\ Rev.\ Lett.\  {\bf 74} (1995) 3327

\bibitem{nrqcd} 
G.~T.~Bodwin, E.~Braaten and G.~P.~Lepage,
Phys.\ Rev.\ D {\bf 51} (1995) 1125
[Erratum-ibid.\ D {\bf 55} (1997) 5853]

\bibitem{cacciari:yr96}
M. Cacciari and M. Kr\"amer, \prl{\bf 76} (1996) 4128.

\bibitem{Amundson:yr97} 
J. Amundson, S. Fleming, and I. Maksymyk, \prd{\bf 56} (1997) 5844.

\bibitem{ko:yr96}
P. Ko, J. Lee, and H.S. Song, \prd{\bf 54} (1996) 4312; [{\bf 60}
(1999) 119902(E)].

\bibitem{Kniehl:yr97}
B.A. Kniehl and G. Kramer, \plb{\bf 413} (1997) 416.

\bibitem{kramer:yr96}
M. Kr\"amer, \npb{\bf 459} (1996) 3.
M. Kr\"amer, J. Zunft, J. Steegborn and P. M. Zerwas, \plb{\bf 348} (1995).

\bibitem{beneke:96yr}
M. Beneke and I.Z. Rothstein, \plb{\bf 372} (1996) 157 
[Erratum-ibid. B{\bf 389} (1996) 769];
M. Beneke and M. Kr\"amer, \prd{\bf 55} (1997) 5269.

\bibitem{braaten:99yr} E. Braaten, B.A. Kniehl, and J. Lee,
\prd{\bf 62} (2000) 094005;
B. A. Kniehl and J. Lee, \prd{\bf 62} (2000) 114027.

\bibitem{leibovich:97yr} Adam K. Leibovich, \prd{\bf 56} 
(1997) 4412.

\bibitem{BABAR1}
B.~Aubert {\it et al.}  [BABAR Collaboration],
Phys.\ Rev.\ Lett.\  {\bf 87} (2001) 162002

\bibitem{BELLE1}
K.~Abe {\it et al.}  [BELLE Collaboration],
Phys.\ Rev.\ Lett.\  {\bf 88} (2002) 052001

\bibitem{hera:h1}
H1 Collaboration, Contributed Paper 157aj, International Europhysics 
Conference on High Energy Physics (EPS99), Tampere, Finland, 1999. 

\bibitem{hera:zeus}
ZEUS Collaboration, Contributed Paper 851, International 
Conference on High Energy Physics (ICHEP2000), Osaka, Japan, 2000.

\bibitem{klasen:yr02}
M. Klasen, B.A. Kniehl, L.N. Mihaila, and M. Steinhauser,
\prl{\bf 89} (2002) 032001.

\bibitem{delphi:yr01}
S. Todorova-Nova, in Proceedings of the 31st International 
Symposium on Multiparticle Dynamics, 
Datong, China, 2001
[ArXiv: hep-ph/0112050];
M. Chapkin, talk at 9th International Conference on Hadron 
Spectroscopy,  Protvino, Russia, 2001, 
AIP Conf. Proc. 619 (2002) 803,
hep-ph/0307049(to appear in PLB).

\bibitem{qiaow0301} 
C. F. Qiao and J. X. Wang, to be appear in \prd.  
[ArXiv: hep-ph/0308244]

\bibitem{qcf:03} 
C.~F.~Qiao, J. Phys. G: Nucl. Part. Phys. 29 (2003) 1075

\bibitem{braaten:96}
E. Braaten and Y. Q. chen, 
Phys.\ Rev.\ Lett.\  {\bf 76} (1996) 730

\bibitem{p.cho:96}
Peter  Cho, Adam K. Leibovich,
Phys.\ Rev.\ D {\bf 53} (1996) 6203
Phys.\ Rev.\ D {\bf 54} (1996) 6690

\bibitem{f.yuan:97}
F. Yuan, C.-F. Qiao, and K.T. Chao,
Phys.\ Rev.\ D {\bf 56}, (1997) 321.
K.T. Chao and L. K. Hao, \npb{\bf 115} (2003) 162.

\bibitem{s.baek:98}
S.~Baek, P.~Ko, J.~Lee and H.~S.~Song,
J.\ Korean Phys.\ Soc.\  {\bf 33} (1998) 97

\bibitem{schuler:99}
G. A. Schuler. Eur. Phys. J. C8,273(1999)

\bibitem{fleming2003} 
S.~Fleming, A. K. Leibovich and T. Mehen.
[arXiv:hep-ph/0306139].

\bibitem{chang:yr97} C.H. Chang, C.F. Qiao, and J.X. Wang,
\prd{\bf 56} (1997) 1363; {\bf 57} (1998) 4035.

\bibitem{BELLE2}
K.~Abe {\it et al.}  [Belle Collaboration],
Phys.\ Rev.\ Lett.\  {\bf 89}, (2002) 142001.

\bibitem{braaten:03}
G. T. Bodwin, J. Lee and E. Braaten
Phys.\ Rev.\ Lett.\  {\bf 90} (2003) 162001

\bibitem{IRC}
A. B. Arbuzov, V. A. Astakhov, A. V. Fedorov, G. V. Fedotovich and E. A. Kuraev
JHEP 9710:006,1997, [ArXiv: hep-ph/9703456]

\bibitem{wang:fdc96} J. X. Wang, FDC project was started at 1992 
and many new parts added since then. 
"Proceeding of AI93, Oberamergau, Germany.
New computing techniques in physics research III" 517-522,"
Appi 2001, Accelerator and particle physics" 108-121.                                                                                
\end{thebibliography}
\end{document}